# First Generation Heterodyne Instrumentation Concepts for the Atacama Large Aperture Submillimeter Telescope

Christopher Groppi, Andrey Baryshev, Urs Graf, Martina Wiedner, Pamela Klaassen, Tony Mroczkowski

*Abstract*— The Atacama Large Aperture Submillimeter Telescope (AtLAST) project aims to build a 50-meter-class submm telescope with >1-degree field of view, high in the Atacama Desert, providing fast and detailed mapping of the mm/submm sky. It will thus serve as a strong complement to existing facilities such as the Atacama Large Millimeter/Submillimeter Array (ALMA).

ALMA is currently the most sensitive observatory covering the atmospheric windows from centimeter through submillimeter wavelengths. It is a very powerful instrument for observing sub-arcminute-scale structures at high, sub-arcsecond spatial resolution. Yet its small field of view (< 15" at 350 GHz) limits its mapping speed for large surveys. In general, a single dish with a large field of view can host large multi-element instruments that can more efficiently map large portions of the sky than an interferometer, where correlator resources and the smaller fields of view of the antennas tend to limit the instantaneous number of beams any instrument can have on the sky. Small aperture survey instruments (typically much smaller than < 3 × the size of an interferometric array element) can mitigate this somewhat but lack the resolution for accurate recovery of source location and have small collecting areas. Furthermore, small aperture survey instruments do not provide sufficient overlap in the spatial scales they sample to provide a complete reconstruction of extended sources (i.e. the zero-spacing information is incomplete in u,v-space.)

Heterodyne instrumentation for the AtLAST telescope will take advantage of extensive developments in the past decade improving the performance and pixel count of heterodyne focal plane arrays. The current state of the art in heterodyne arrays are the 64-pixel Supercam instrument, the 16-pixel HARP instrument, the dual band SMART receiver with 8 pixels in two bands, and the GREAT instrument on SOFIA with 21 pixels (14 at 1.9 THz and 7 at 4.7 THz). Future receivers with larger pixel counts have been under development: CHAI for CCAT (64-pixels) and SHASTA for SOFIA (64 pixels) or under study, e.g. HERO for the Origins Space Telescope (2x9 to 2x64 pixels). Instruments with higher pixel counts have begun to take advantage of integration in the focal planes to increase packaging efficiency over simply stacking modular mixer blocks in the focal plane.

The authors believe that heterodyne instruments with pixel counts of approximately 1000 pixels per band could be considered for AtLAST on a decade timescale. The primary limiting factor in instrument capability (pixel count, instantaneous bandwidth, number of frequency bands, polarization capability, side-band separation etc.) is likely to be cost, rather than any fundamental technological limitation. As pixel counts increase, the cost and complexity of the IF system and spectrometer also rapidly increases, particularly if wide IF bandwidth, dual polarization and sideband separation is desired. Currently the IF and backend are limited by the cost and power consumption per unit bandwidth of the total processed science signal. While that cost is likely to decrease modestly in the next decade, no technology is likely to disrupt the scaling argument. Many of the front-end costs will also scale with pixel count, for example the size and cooling capacity of the cryostat, the complexity of the LO subsystem, and the I&T cost associated with developing, assembling and testing the focal plane units.

In this presentation, we review the state of the art for millimeter/sub-millimeter heterodyne instrumentation technology that could be suitable for AtLAST and attempt to forecast how the technologies will advance over the next decade. We then present a design concept for a potential first-generation AtLAST heterodyne instrument. These considerations meet the scientific demands and atmospheric considerations for a ground-based facility in the Atacama Desert.

## I. Introduction

THE Atacama Large Millimeter/Submillimeter Array (ALMA) is currently the most sensitive observatory that can cover the atmospheric windows from millimeter through submillimeter wavelengths (35-950 GHz). Yet its small field of view (<18" at 850 microns) limits its mapping speed for large surveys. In general, a large single dish with a large field of view can host large multi-element instruments that can more

Manuscript submitted June 25, 2019. The project leading to this publication has received funding from the European Union's Horizon 2020 research and innovation program under grant agreement No 730562 [RadioNet].

C.E. Groppi is with the School of Earth and Space Exploration, Arizona State University, Tempe, AZ 85287 USA (e-mail: cgroppi@asu.edu).

A.M. Baryshev is with the Kapteyn Astronomical Institute, University of Groningen, Groningen, 9700 AB NL (e-mail: andrey@astro.rug.nl).

U.U. Graf is with the Institute of Physics, University of Cologne, Cologne, 50937 Germany (e-mail: graf@ph1.uni-koeln.de).

M.C. Wiedner is with Sorbonne Université, Observatoire de Paris, Université PSL, CNRS, 75014 Paris, France (e-mail: martina.wiedner@obspm.fr).

P. Klaassen is with the UKATC, Royal Observatory Edinburgh, Edinburgh, Scottland, UK EH9 3HJ (email: pamela.klaassen@stfc.ac.uk).

T. Mroczkowski is with the European Southern Observatory, Garching, 85748 Germany (e-mail: tony.mroczkowski@eso.org).

efficiently map large portions of the sky than an interferometer, where correlator resources and the smaller fields of view of the array elements tend to limit the instantaneous number of beams the instrument has on sky.

Small aperture (6-meter) survey instruments like CCAT-prime [1] can mitigate this somewhat but lack the resolution for accurate recovery of source location. Furthermore, small aperture survey instruments do not provide sufficient overlap in the spatial scales they sample to provide a complete reconstruction of extended sources (i.e. the zero-spacing information is incomplete in u,v-space).

The Atacama Large Aperture Submillimeter Telescope (AtLAST) project aims to fill this technological capability gap (see [1]). AtLAST is a (domeless) 50-meter class dish with surface accuracy sufficient to provide good aperture efficiencies up to at least 950 GHz, covering the 350-micron window crucial for studies of both local and high-redshift star formation. It will feature a large field of view (> 1 degree, i.e. a factor of 225 times that of the 50-meter Large Millimeter Telescope) and a receiver cabin sufficiently large to host a broad suite of heterodyne and direct-detection instrumentation. AtLAST will strongly complement the high-resolution follow-up capabilities of ALMA, while delivering unique survey and targeted capabilities.

Basic considerations for the site were presented in [2] and include sites ranging from the planes to the peak of Chajnantor; i.e. this includes Llano de Chajnantor at 5100 meters above sea level, up to 5600 meters at Cerro Chajnantor. Upon first light, AtLAST will be fully outfitted with a number of instruments providing complementary capabilities such as broad instantaneous bandwidth, widefield survey capabilities, and the ability to explore new wavelength regimes. In this work, we discuss the state of the art for several heterodyne millimeter/sub-millimeter instrumentation technologies that could populate AtLAST's >1 square degree focal plane, and attempt to forecast how the technologies will advance over the next decade. Some factors considered are the bandwidth, spectral resolution, and multi-element capabilities, and how these couple to the AtLAST design concept advanced in the telescope design report. We then present a few on-paper design concepts for potential first-generation heterodyne instruments for AtLAST. These considerations are informed both the scientific demands and atmospheric considerations for a ground-based facility high in the Atacama Desert.

The past two decades have seen extensive surveys of the far-infrared to submillimeter continuum emission in the plane of our Galaxy. We line out prospects for the coming decade for corresponding molecular and atomic line surveys which are needed to fully understand the formation of the dense structures that give birth to clusters and stars out of the diffuse interstellar medium. We propose to work towards Galaxy wide surveys in mid-J CO lines to trace shocks from colliding clouds, Galaxy-wide surveys for atomic Carbon lines in order to get a detailed understanding of the relation of atomic and molecular gas in clouds, and to perform extensive surveys of the structure of the dense parts of molecular clouds to understand the importance of filaments/fibers over the full range of Galactic environments and to study how dense cloud cores are formed from the filaments. This work will require a large (50 m) Single Dish submillimeter telescope equipped with massively multipixel spectrometer arrays, such as envisaged by the AtLAST project.

## II. POTENTIAL FIRST-GENERATION HETERODYNE INSTRUMENTS

Heterodyne instrumentation for the AtLAST telescope will take advantage of extensive developments in the past decade improving the performance and pixel count of heterodyne focal plane arrays [4]. The current state of the art in heterodyne arrays are the 64-pixel Supercam instrument [5], the 16-pixel HARP instrument [6], the dual band SMART receiver with 8 pixels in two bands [7], and the GREAT instrument on SOFIA with 21 pixels (14 at 1.9 THz and 7 at 4.7 THz) [8]. Future receivers with larger pixel counts have been under development: CHAI for CCAT (64-pixels) and SHASTA for SOFIA (64 pixels). Instruments with higher pixel counts have begun to take advantage of integration in the focal planes to increase packaging efficiency over simply stacking modular mixer blocks in the focal plane. The AtLAST heterodyne instrument group believes that instruments with pixel counts as high as 1000 pixels could be considered for AtLAST on a decade timescale. The primary limiting factor in instrument capability (pixel count, instantaneous bandwidth, number of frequency bands, polarization capability, sideband separation etc.) is likely to be cost rather than any fundamental physical or engineering limit. As pixel counts increase, the cost and complexity of the IF system and spectrometer also rapidly increases, particularly if wide IF bandwidth, dual polarization and sideband separation is desired. Currently the IF and backend are limited by the cost per unit bandwidth of the total processed science signal. While that cost is likely to decrease modestly in the next decade, no technology is likely to disrupt the scaling argument. Many of the front-end costs will also scale with pixel count, for example the size and cooling capacity of the cryostat, the complexity of the LO subsystem, and the I&T cost associated with developing, assembling and testing the focal plane units.

Two areas where fundamental work is still required are mixer sensitivity and array packaging of complex mixer topologies. SIS receivers are reaching the few h/k noise level with the development from projects like ALMA and Herschel. But, packaging mixers into arrays results in unavoidable compromises due to size constraints in array packaging, widefield optics and relatively large cryostat windows and IR filters required. Further work is required to reduce the noise of pixel elements in large format arrays to the current state of the art or better for single pixel receivers (e.g. ALMA & Herschel) to fully take advantage of the expected AtLAST site.

In addition, more work is needed to efficiently package complex mixer topologies (e.g. 2SB mixers) in large format focal plane arrays. All the arrays mentioned above reach their pixel counts using relatively simple dual sideband mixers combined with quasi-optical LO injection, polarization diplexing and band separation. Heterodyne focal plane arrays with OMT based polarization diplexing and balanced, single sideband or sideband separating capabilities have yet to be demonstrated.

The AtLAST heterodyne instrument group also recommends close collaboration at this point in the project as requirements are generated for science instruments. Due to the cost limitation we expect to come into play, careful optimization of requirements is necessary to extract the maximum science from a future AtLAST heterodyne array. Participation of instrumentation specialists would be highly beneficial in optimizing the science vs. cost trade space for such an instrument.

*A. Field of View*

The number of heterodyne pixels that can be accommodated in the telescope focal plane grows with the square of the telescope diameter and the square of the field of view diameter:

$$N \sim 2 \times 10^5 \left(FOV/1° \times D/25m \times 350\mu m/\lambda\right)^2 \quad (1)$$

Several hundred thousand pixels would fit in the focal plane of AtLAST, but due to the complexity and cost of each pixel, it does not seem realistic to fully populate the focal plane on the timescale for AtLAST first light instrumentation. With current technology the cost of the IF chain for each channel from the low noise amplifier to the spectrometer is on the order of 25,000 US-$ or more. Quite likely this cost can be reduced if large quantities are produced, but we do not expect to see an order of magnitude price change. Together with the cost for common components like local oscillators, cryostats and refrigerators, a 1000-pixel array will likely require several tens of millions US-$. In the foreseeable future financial constraints will limit heterodyne arrays to approximately 1000 pixels, and thereby only filling ~1% of the focal plane of a large telescope. With this limitation, the science trades required are to determine how to allocate these pixels. The cost of an instrument with four 256 pixel frequency bands is of the same order of magnitude as a single band with 1000 pixels. The added optics and front-end complexity of a four-band instrument will be offset by savings realized in sharing IF and backend hardware, at least at the rough order of magnitude level presented here.

*B. 2SB vs. DSB for ground-based arrays*

Sideband separating (2SB) heterodyne receiver pixels are more complex than DSB pixels but have several key advantages. First, they allow for very accurate sideband ratio calibration because both the LSB and USB are available at different outputs, second 2SB receiver rejects atmospheric noise from the image sideband improving system noise temperature and the scanning speed.
Typically, a 2SB array pixel will have one input horn, two mixers, two low noise amplifiers and IF and RF hybrid circuits. In practice this is as complex and as expensive as two DSB array pixels. Therefore, 2SB pixels are only efficient to implement in an FPA if it they offer at least double the scanning speed relative to DSB pixels.

The scanning speed $V_s$ of heterodyne system can be expressed as:

$$V_s = \frac{\Delta f \, \Delta T^2}{T_{sys}^2}, \quad (2)$$

where $V_s$ is the frequency independent scanning speed in receiver beams per second, $\Delta f$ is the channel bandwidth, $\Delta T$ is the required brightness noise level in K, and $T_{sys}$ is the system noise temperature including the contribution of both the receiver and atmosphere. One must note that scanning speed depends on $T_{sys}^2$ which is a strong dependence. Special attention should be paid not to degrade the $T_{sys}$ of a single pixel compared to its performance in the array to realize the benefit in scanning speed.

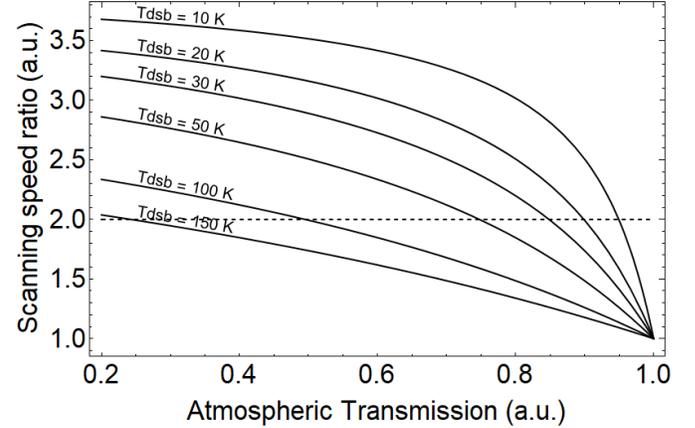

Fig. 1. Scanning speed ratio $R_{2SB/DSB}$ for typical atmospheric temperature $T_{atm} = 280\,K$ and different atmospheric transmission $\tau$ values and DSB receiver noise temperatures $T_{dsb}$. Threshold value where 2SB pixel is twice faster is indicated by the dashed line.

System noise temperature of a ground-based receiver can be calculated using following equation:

$$T_{sys} = 2\frac{T_{dsb}}{\tau} + \frac{T_{atm}(1-\tau)(1+SBR)}{\tau}, \quad (3)$$

where $T_{dsb}$ is the receiver equivalent DSB noise temperature, $T_{atm} = $ is the physical temperature of the lower layer of atmosphere, $T_{dsb}$ is the receiver DSB noise temperature, $\tau$ is atmospheric transmission and $SBR$ is the sideband ratio which is the ratio of receiver gain in the image sideband to the receiver gain in the signal sideband. For an ideal DSB receiver, the $SBR$ is equal 1 and for a 2SB receiver is in the range of 0.1 to 0.01 [refs].

Following equations (2) and (3) the scanning speed ratio $R_{2SB/DSB}$ of 2SB over DSB mixers can be expressed as:

$$R_{2SB/DSB} = \left(\frac{2\,T_{dsb} + 2\,T_{atm}(1-\tau)}{2\,T_{dsb} + T_{atm}(1-\tau)(1+SBR)}\right)^2. \quad (4)$$

For an ideal 2SB receiver, ($SBR=0$) this depends only on the atmospheric brightness: ratio, $T_{br}/T_{dsb}$:

$$R_{2SB/SSB} = \left(\frac{2 + 2\,T_{br}/T_{dsb}}{2 + T_{br}/T_{dsb}}\right)^2, \quad (5)$$

where $T_{br} = T_{atm}(1-\tau)$ is atmospheric noise seen by a ground-based receiver which represents the background limit for given atmospheric conditions. Without the presence of the atmosphere the $R_{2SB/SSB}$ equals unity and DSB and 2SB pixels are equivalent to each other. It is the influence of atmospheric noise makes a 2SB array pixel more efficient. With a fully background limited receiver $T_{br} \gg T_{DSB}$ the scanning speed of 2SB pixel can be significantly larger than a DSB pixel. Since a 2SB array pixel involves two DSB mixers, amplifiers and RF and IF hybrids, we estimate that it is approximately equivalent to two DSB pixels in terms of complexity and cost. Therefore we consider $R_{2SB/SSB} = 2$ as threshold value where 2SB pixel is equivalent to 2 DSB pixels in scanning speed, which corresponds to $T_{br}/T_{dsb} = \sqrt{2}$.

In reality, the $R_{2SB/SSB}$ depends on atmospheric conditions and DSB noise temperature of mixers as shown in figure 2. If $T_{dsb}$ is small, a 2SB pixel will significantly outperform 2xDSB pixels under large range of atmospheric transmission.

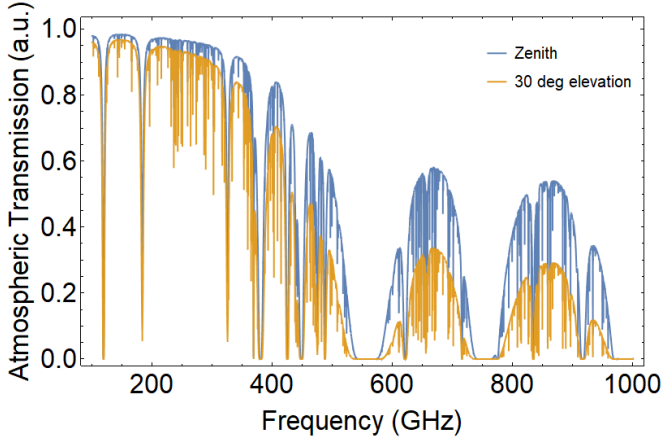

Fig. 2. Zenith and 30 Deg elevation atmospheric transmission $\tau$ for ALMA site at Chajnantor Plato, Atacama Desert in Chile. Values represents 25th percentile.

Let us consider atmospheric conditions on ALMA site shown in figure 2 for the 25th percentile. Atmospheric transmission is presented for zenith and 30 deg elevation which contains all typical observing elevations. The typical atmospheric brightness at ALMA site derived from fig. 2 is shown in fig. 3 in comparison with the average value of ALMA receiver band noise temperatures. While the most ALMA receivers are of 2SB type we present the equivalent DSB noise temperature which is half of the SSB noise temperature and is directly applicable in equations (1-4). Noise temperatures in units of the quantum limit are also shown for comparison. From this data, we can conclude that significant part of observing time even at the best available submm site, the system noise is dominated by atmosphere and thus the 2SB pixels are clearly beneficial over DSB for frequencies above 200 GHz.

### III. POTENTIAL INSTRUMENT CONFIGURATIONS

Section II.B makes a strong argument for the use of 2SB mixer pixels as long as their complexity in an array configuration can be handled (e.g. with on chip SSB technology). It is then feasible to consider ~500 SSB pixels given the cost constraints of a ~$25M USD instrument budget. How these pixels should be configured is then the main trade. Pixels configured as a dual polarization system adds some complexity to the design but is more well suited to making deeper, small area maps (e.g. for imaging of resolved extragalactic sources). For widefield mapping of the Galaxy, the mapping speed is identical for pixels all in one polarization or split between two polarizations. We will consider two possible configurations: a 512-pixel 2SB array covering ALMA band 6 implemented as a pair of 256 pixel sub-arrays in a dual polarization configuration, and four 128-pixel arrays covering four ALMA bands from bands 6 to 10 implemented as a pair of 64 pixel sub-arrays also in a dual polarization configuration.

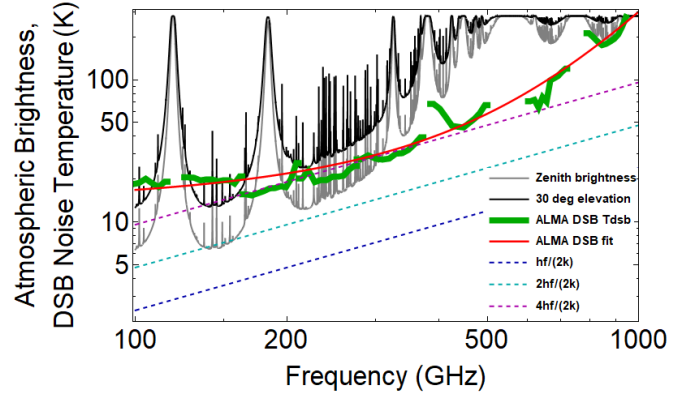

Fig. 3. Atmospheric brightness compared to equivalent DSB receiver noise temperature of ALMA receivers and quantum limit.

#### A. Large single band array

For the purposes of a large Galactic plane survey with the goal of mapping the maximum area as fast as possible, a single band instrument operating at ALMA band 6 would be the most ideal choice. Mapping speed scales linearly with the number of pixels and is inversely proportional to the square of the system temperature. This favors a large number of pixels at a lower frequency band where both the receivers and the sky are the best. In addition, the field of view covered by the array scales inversely with the square of the frequency of operation simply because of the change in the diffraction limited beam size. This also strongly favors lower frequencies for maximum mapping speed. Assuming $T_{rec}=3h\nu/k$, atmospheric transmission of 0.95 (figure 2), and a 2SB scanning speed ratio of 3, such an instrument could map to a depth of 1 mJy in 100 kHz channels at a speed of approximately 30 hours per square degree not including overheads. Meeting this receiver temperature goal combined with 2SB mixer pixels in an array with 512 pixels still requires significant future technology development but is not unreasonable to consider on a decade timescale.

#### B. Smaller arrays in multiple bands

While maximum mapping speed will be realized by implementing all pixels in one band, the scientific flexibility of smaller arrays covering multiple bands would be more useful for a larger number of projects. We estimate that the cost of implementing four 128-pixel arrays in four bands to be approximately the same (within a factor of two) as the single larger array covered in section III.A. We expect the cost of the multiple cryostats, additional integration and test and band selection optics to be offset by the savings realized in a smaller backend shared between the four bands. Such an instrument could cover four ALMA like bands between 230 GHz and 950 GHz, selected to maximize science output. Such a configuration would also allow relatively cost-effective incremental upgrades to add additional frequency bands as funds become available.

The 230 GHz channel would map at a speed four times slower than the instrument in section III.A, 120 hours per square degree to 1 mJy in 100 kHz channels. Mapping speeds of higher frequency bands would be significantly slower, scaling like $(\nu*T_{sys})^{-2}$. Assuming receiver pixels still can be developed with $T_{rec}=3h\nu/k$ and the sky noise contribution scales roughly linearly with frequency, the mapping speed of higher frequency bands to a given depth unavoidably drops like $\nu^4$. This drop can only be mitigated by increasing the number of pixels at higher frequencies.

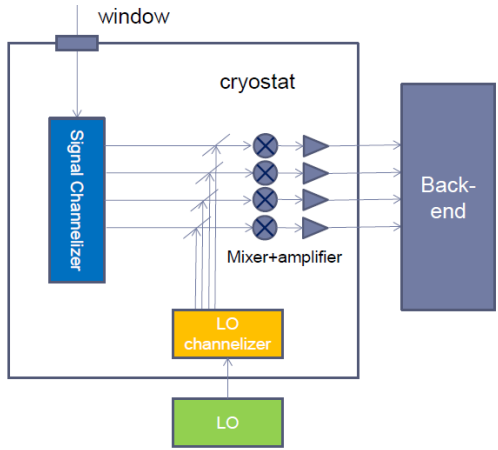

Fig. 4. Principle diagram of a quasi-optically combined frequency array receiver.

## IV. FREQUENCY ARRAY RECEIVER CONCEPT

The concept of combining several spatial pixels into a focal plane array can be naturally extended in arranging pixels not in focal plane of a telescope, but in frequency space as presented in figures 4 and 5 resulting in a frequency array. The telescope signal from one spatial pixel on the sky is split into narrow frequency sub-bands using either quasioptical or planar filter. Each sub-band then feeds a SIS mixer. A comb of equally spaced LO signals are coupled to each of the mixer in such a way that continuous and simultaneous RF frequency coverage is achieved. The IFs of each frequency pixel are amplified and analyzed with the same type of back-end as for a conventional FPA. The input channelizing filter bandwidth of such a receiver should be matched with each pixel IF bandwidth. For current state of art mixers, this can be as wide as 16 GHz or more

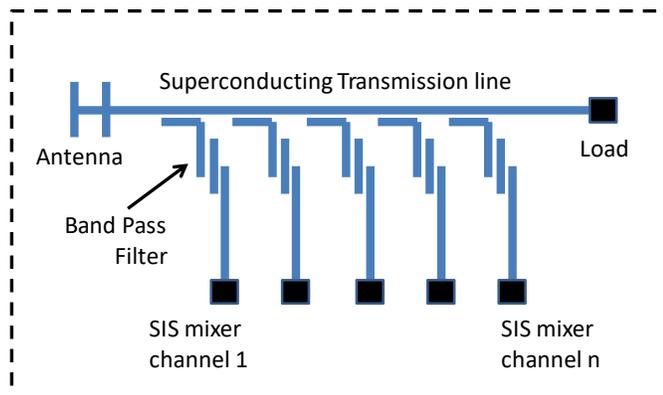

Fig. 5. Principle diagram of an on-chop combined frequency array receiver.

[9],[10]. In addition to quasi-optical free space filters, the channelization can be achieved in a very compact way on-chip as shown in figure 5. This concept is similar to filter bank on chip [11], [12] with the notable difference that the filter channel bandwidth is much larger, with high spectral resolution achieved through the use of a heterodyne receiver at the output of each channelizing filter. The frequency coverage of the frequency array is shown schematically in figure 6. In comparison with a direct detection receiver, the frequency information of incoming photons is retained in heterodyne reception. This design avoids efficiency loss due to channelization filter overlap, as the signal from neighboring bands can be recombined if the individual instantaneous coverage of each heterodyne mixer is slightly larger than 3dB bandwidth of each channel, i.e. the mixer IF bands have small frequency overlaps. The same argument allows significant relaxation of the requirement for filter edge sharpness, while maintaining high, close to unity overall coupling from input beam to each frequency pixel.

### A. Sensitivity and utilization

The frequency array receiver can be designed to have an extremely large instantaneous bandwidth. A 600 GHz band can be covered by a rather modest 40 frequency pixels of 16 GHz IF bandwidth. This bandwidth is more than 50 times larger than typical ALMA IF coverage, which means that for blind frequency searches of red shifted transmission lines this system can have the same speed as whole ALMA array if implemented on a 12m diameter dish. On AtLAST, the point source sensitivity would be increased by another factor of 4-17.

For total power observations, the total bandwidth of frequency array receiver is much larger than any direct detecting system, which covers typically only one atmospheric window while avoiding prominent atmospheric absorption lines. As illustrated by figure 2, the density of telluric absorption lines is significant, and their additional background contribution cannot be avoided by quasioptical filters. The frequency array receiver noise is not influenced by these lines and the spectral resolution is sufficient to resolve the lines and exclude the high background channels while still leaving large equivalent bandwidth for sensitive observation.

In comparison with direct detector systems, the quantum limit of heterodyne receivers presents significant limitation for low background systems. Direct detectors have no sensitivity limit and can be designed to have superior noise performance. However, as illustrated in figure 3, state of the art heterodyne ground-based receivers are sky background limited for all submm bands. Given the other advantages like bandwidth and frequency resolution, the frequency array receiver will be superior to both on-chip filter bank spectrometers, grating spectrometer and total power spatial pixels for ground-based applications.

Finally, when covering the full THz receiver window (100-1000GHz) the frequency array receiver will be the ultimate receiver for an radio interferometer, like ALMA or SMA as it will provide all complex signal information over the full ground based sub-mm band instantaneously, providing ultimate flexibility and sensitivity if such bandwidths are supported by a future correlator.

## V. Conclusion

Technology developments in the past decades make large format (~1000 pixel) heterodyne arrays with near quantum and background limited noise performance possible for telescopes like AtLAST. The primary limitation that will drive engineering vs. science output decisions is cost rather than technical readiness. We believe reasonable cost limitations on the order of tens of millions of USD to be dedicated to heterodyne instrumentation on AtLAST will limit the number of pixels to ~1000. These pixels can be deployed in many ways, from a single band large format array to yield the maximum possible mapping speed, or in smaller arrays covering multiple bands. 2SB mixer topologies have the potential to significantly improve performance for such receivers but do come at a cost increased complexity.

A novel frequency array receiver concept has been presented that could allow coverage of the entire sub-mm band instantaneously with heterodyne resolution. Such a receiver would be synergistic with a large focal plane array, allowing very deep and wideband point source observations. This instrument could also share much of the backend system required for a focal plane array, thereby reducing the cost of implementation.

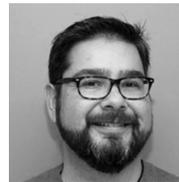

**Christopher E. Groppi** received the B.A. degree in astronomy (with honors) from Cornell University in Ithaca, NY, in 1997 and the Ph.D. degree in astronomy with a minor in electrical and computer engineering from the University of Arizona in Tucson, AZ, in 2003.

In 2003, he joined the National Radio Astronomy Observatory as a Director's Postdoctoral Fellow. He then moved to the University of Arizona as an assistant staff astronomer in 2005. In 2006, he received an Astronomy and Astrophysics Postdoctoral Fellowship from the National Science Foundation. In 2009, he joined the Arizona State University School of Earth and Space Exploration in Tempe, AZ as an assistant professor. He became an associate professor in 2015. He is an experimental astrophysicist interested in the process of star and planet formation and the evolution and structure of the interstellar medium. His current research focuses on the design and construction of state of the art terahertz receiver systems optimized to detect the light emitted by molecules and atoms in molecular clouds, the birthplace of stars. He also applies terahertz technology developed for astrophysics to a wide range of other applications including Earth and planetary science remote sensing, hazardous materials detection and applied physics.

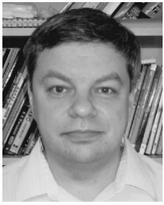
**Andrey M. Baryshev** received the M.S. degree (summa cum laude) in physical quantum electronics from the Moscow Physical Technical institute, Moscow, Russia, in 1993, and the Ph.D. degree in superconducting integrated receiver combining SIS mixer and flux flow oscillatoron into one chip from the Technical University of Delft, Delft, The Netherlands, in 2005.

He is currently an associate professor at the Kapteyn Astronomical Institute, University of Groningen, Groningen, and was previously a Senior Instrument Scientist with the SRON Low Energy Astrophysics Division, Groningen, The Netherlands. In 1993, he was an Instrument Scientist with the Institute of Radio Engineering and Electronics, Moscow, involved in the field of sensitive superconducting heterodyne detectors. In 2000, he joined an effort to develop an SIS receiver (600–720 GHz) for the atacama large millimiter array, where he designed the SIS mixer, quasi-optical system, and contributed to a system design.

His current main research interests include application heterodyne and direct detectors for large focal plane arrays in THz frequencies and quasi-optical systems design and experimental verification.

Dr. Baryshev was the recipient of the NWO-VENI Grant for his research on heterodyne focal plane arrays technology in 2008 and, in 2009, he was the recipient of the EU commission Starting Researcher Grant for his research on focal plane arrays of direct detectors.

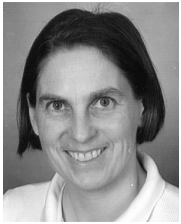
**Martina C. Wiedner** received the Ph.D. degree in physics from the University of Cambridge, Cambridge, U.K., in 1999.

She was a Postdoctoral Fellow with the Harvard Smithsonian Center for Astrophysics, Boston, MA, USA, and she also led a junior research group with the University of Cologne, from 2003 to 2009. Since 2009, she has been a Permanent Researcher with the Laboratoire d'Etude du Rayonnement et de la Matiere en Astrophysique, Observatory of Paris, ´Paris, France. Her main research is focused on the development of heterodyne instrumentation for astronomy and earth sciences.

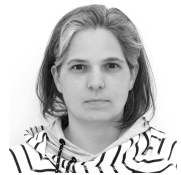
**Pamela D. Klaassen** received a B.Sc and M.Sc in astronomy from the University of Calgary in 2002 and 2004 respectively. She then moved to McMaster University in Hamilton, Ontario, Canada for her Ph.D. which she obtained in 2008. During that time, she also spent a year at the Harvard-Smithsonian Center for Astrophysics as an SMA pre-doctoral fellow.

She did her post-doctoral fellowships in the European ALMA regional Centres at ESO in Garching, Germany (2008-2011) and Leiden Observatory, in Leiden, the Netherlands (2011-2014). Since then, she has been an Instrument Scientist at the UK Astronomy Technology Centre, where she is the science lead for the JWST/MIRI simulator, and for the team developing the Observatory Science Operations software for the SKA. She is the lead of the science working groups for the AtlAST project, and her scientific interests focus on the formation of the most massive stars in the Milky Way.

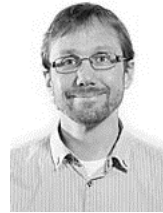
**Tony Mroczkowski** received his B.S. in engineering from the Cooper Union for the Advancement of Science and Art in 2001 and his Ph.D. in astronomy from Columbia University in New York City, NY in 2009 (defended in 2008). From 2008-2011, Mroczkowski was a postdoctoral researcher at the University of Pennsylvania in Philadelphia, PA, the last year of which was funded by the NASA Einstein Fellowship program. Mroczkowski continued this fellowship at NASA Jet Propulsion Laboratory in Pasadena, CA, in 2011-2013. Mroczkowski was then awarded the National Research Council Fellowship to work at the Naval Research Laboratory in Washington, D.C. from 2013-2016. In 2016, he became assistant level faculty at the European Southern Observatory in Garching, Germany, working as an astronomer / (sub)millimeter instrument scientist on ALMA development. He is an experimental cosmologist primarily interested in the Sunyaev-Zel'dovich effect from galaxy clusters, groups, and large scale structure.